\documentclass[sigconf]{acmart}
\usepackage[T1]{fontenc}
\usepackage{newtxtext}
\usepackage{xspace}
\usepackage{csquotes}
\usepackage[noorphans,vskip=1em,leftmargin=1em]{quoting}
\usepackage[normalem]{ulem}
\usepackage{algorithm}
\usepackage{algpseudocode}
\usepackage{hhline}
\usepackage{subfig}
\usepackage[table]{xcolor}
\definecolor{fm1}{HTML}{AA2B46}
\definecolor{fm2}{HTML}{C65F10}
\definecolor{fm3}{HTML}{E37400}
\definecolor{fm4}{HTML}{174EA6}
\usepackage[most]{tcolorbox}
\usepackage{csquotes}
\usepackage[noorphans,vskip=1em,leftmargin=1em]{quoting}

\usepackage{amsmath}

\usepackage{booktabs}
\usepackage{multirow}
\usepackage{makecell}
\usepackage{threeparttable} 
\usepackage{xcolor}
\usepackage{graphicx} 
\definecolor{ForestGreen}{RGB}{34,139,34}
\definecolor{grey}{gray}{0.95}

\usepackage{array}
\usepackage{siunitx}
\usepackage{enumitem}

\definecolor{DeepRed}{RGB}{196,59,59}
\definecolor{LightRed}{RGB}{252,239,239}

\setlist[itemize]{
    labelindent=2pt,
    leftmargin=*,
    labelsep=0.5em
}

\newcommand{\approach}{E\textsc{nv}G\textsc{raph}\xspace}

\definecolor{AlgoComment}{RGB}{170,60,130}   
\definecolor{AlgoPhase}{RGB}{170,0,85}     
\definecolor{AlgoGray}{RGB}{90,90,90}

\algrenewcommand\algorithmicrequire{\textbf{Input:}}
\algrenewcommand\algorithmicensure{\textbf{Output:}}

\algrenewcommand\alglinenumber[1]{\footnotesize\bfseries #1}

\algrenewcommand\algorithmiccomment[1]{\hfill{\footnotesize\textcolor{AlgoComment}{\#~#1}}}

\newcommand{\AlgHeader}[1]{\Statex \hspace{-\algorithmicindent}#1}

\newcommand{\AlgPhase}[1]{\Statex \textcolor{AlgoPhase}{\#~#1}}
\definecolor{oai-orange}{HTML}{FE7600}

\AtBeginDocument{%
  }

\setcopyright{none}
\newcommand{\inst}[1]{\textsuperscript{#1}}

\author{
Ruwei Pan\inst{1,2} \quad
Junlei Shen\inst{1} \quad
Linhao Wu\inst{2} \quad 
Yueheng Zhu\inst{1,2} \quad
Zixiong Yang\inst{2} \\[2pt]
Yakun Zhang\inst{3} \quad
Lu Zhang\inst{2} \quad
Hongyu Zhang\inst{1}
}

\authornote{
\inst{1} Chongqing University \quad
\inst{2} Peking University \quad
\inst{3} Harbin Institute of Technology (Shenzhen)
}

\begin{document}

\title{Toward Executable Repository-Level Code Generation via Environment Alignment}




\begin{abstract}
Large language models (LLMs) have achieved strong performance on code generation, but existing methods still struggle with repository-level code generation under executable validation. Under this evaluation setting, success is determined not by the plausibility of isolated code fragments, but by whether a generated multi-file repository can be successfully installed, have its dependencies and internal references resolved, be launched, and be validated in a real execution environment. To address this challenge, we propose \approach, a framework for repository-level code generation that formulates repository executability as an environment alignment problem. \approach jointly models two coupled conditions for successful repository execution, namely external dependency satisfaction and repository-internal reference resolution. It maintains a dual-layer environment representation, uses execution evidence to perform execution-evidence-based attribution, and guides repository generation through a unified targeted revision mechanism within an iterative alignment loop. We evaluate \approach on repository-level code generation with three representative backbone LLMs and compare it against representative environment-aware and repository-level baselines. Experimental results show that \approach consistently achieves the best performance on these repository-level benchmarks. 
In particular, it outperforms the strongest non-\approach baseline by an absolute margin of 5.72--5.87 percentage points in Functional Correctness and 4.58--8.66 percentage points in Non-Functional Quality.

\end{abstract}



\maketitle

\section{Introduction}

Large language models (LLMs) have substantially advanced code generation and have achieved strong performance on tasks ranging from function completion to repository-level code completion~\citep{jiang2026survey, liu2023repobench, yang2024execrepobench, li2025fea}. As code generation moves beyond single functions and files toward \emph{repository-level generation}, models are increasingly required to construct complete multi-file repositories from high-level requirements~\citep{ding2025nl2repo}. Under this evaluation setting, the objective is no longer merely to generate plausible code fragments, but to deliver an \emph{executable repository} that can be successfully installed, satisfy its external dependencies, resolve its internal references, be launched, and be validated in a real execution environment. 
Therefore, ensuring repository executability becomes a fundamental prerequisite for repository-level code generation \citep{hai2025impactscontextsrepositorylevelcode}.


However, ensuring repository executability in repository-level generation remains fundamentally challenging~\citep{liu2023repobench, yang2024execrepobench}. 
In repository-level generation, execution failures are not solely caused by logic faults. Many failures instead arise because the generated repository is not yet aligned with the execution conditions required for end-to-end validation. More specifically, successful repository execution depends on two coupled conditions: \emph{external dependency satisfaction} and \emph{repository-internal reference resolution}. \emph{External dependency satisfaction} requires third-party libraries, version constraints, and API migration requirements to be compatible with the target environment~\citep{wu2024versicode, kuang2025apimig}. \emph{Repository-internal reference resolution} requires files, modules, imports, and symbol references within the repository to remain consistently connected and resolvable. As a result, repository execution may fail because the external execution context is unsatisfied, because repository-internal references are broken, or because both conditions are violated. \textbf{The key challenge, therefore, is not merely to observe execution failure, but to determine which failure source is currently dominant under real execution conditions.}


Existing work has addressed this challenge from two main directions. One line of work incorporates external execution constraints into code generation by modeling library versions, API compatibility, and migration requirements~\citep{kuang2025apimig, wu2024versicode, zhang2023repocoder, ding2023crosscodeeval}. These methods improve awareness of external compatibility constraints. However, they are mainly designed to make generated code compatible with external libraries, versions, and APIs. They do not explicitly distinguish whether a surfaced execution failure is caused mainly by unsatisfied external dependencies or by broken repository-internal references. 
Another line of work improves repository-level generation through planning, executable validation, and structural reasoning across files and components~\citep{bairi2024codeplan, ouyang2024repograph, hu2025repo2run}. These methods strengthen repository-level reasoning and often use execution feedback to continue revision. However, they typically treat surfaced execution signals as cues for further revision, rather than as evidence for identifying the currently dominant failure source. Consequently, later revisions may be guided mainly by surfaced execution symptoms rather than by a diagnosis of the dominant source of misalignment, which can lead to misdirected refinement and inefficient iteration.

To address these limitations, we propose \textbf{\approach}, a framework for repository-level code generation that formulates repository generation as an iterative process of environment alignment. The core idea is to jointly model external dependency satisfaction and repository-internal reference resolution, perform attribution based on execution evidence, and guide subsequent generation accordingly. 
Specifically, \approach maintains a dual-layer environment representation to jointly capture the two major sources of repository executability failure. One layer models external dependency satisfaction by capturing dependency requirements, inferred third-party packages, and compatibility constraints exposed during execution. The other models repository-internal reference resolution by capturing repository structure, import and reference relations, and unresolved repository-internal references. Given build results, runtime errors, stack traces, and test outcomes, \approach identifies the dominant source of misalignment and uses this diagnosis to guide the next generation iteration. By continuously updating the environment representation and execution evidence, \approach forms an iterative alignment loop for repository-level code generation.

We evaluate \approach on repository-level code generation across multiple backbone LLMs and compare it against representative environment-aware and repository-level baselines. Results show that \approach consistently improves both functional correctness and non-functional quality under realistic validation settings. In particular, \approach outperforms the strongest non-\approach baseline by 5.72--5.87 points in functional correctness and by 4.58--8.66 points in non-functional quality across different backbone models. Additional ablation and error analyses further reveal the contribution of each component and the remaining limitations of \approach.

In summary, this paper makes the following contributions:
\begin{itemize}
    \item We reformulate repository-level code generation as an \emph{environment alignment} problem and show that repository executability depends on both external dependency satisfaction and repository-internal reference resolution.

    \item We propose \approach, a framework for repository-level code generation that jointly models these two conditions, performs execution-evidence-based attribution to identify the dominant source of misalignment, and improves generation through an iterative alignment loop.

    \item \approach consistently outperforms representative environment-aware and repository-level baselines across multiple backbone LLMs. In particular, it surpasses the strongest non-\approach baseline by 5.72--5.87 points in functional correctness and by 4.58--8.66 points in non-functional quality under realistic validation settings. We further provide detailed ablation studies and error analyses.
\end{itemize}

\section{Motivating Example}

\begin{figure}[t]
    \centering
    \includegraphics[width=0.95\linewidth]{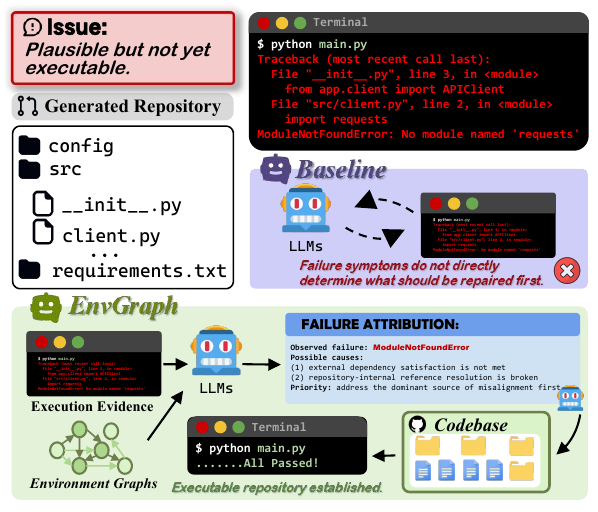}
    
    \caption{A motivating example of repository-level code generation under executable validation. The same execution symptom may arise from unmet external dependencies, broken repository-internal references, or both.}
    \label{fig:motivating_example}
    \vspace{-15px}
\end{figure}

Figure~\ref{fig:motivating_example} shows a motivating example of repository-level code generation under executable validation. The generated repository appears plausible at first glance: it contains multiple files, the main components are present, and the overall structure looks reasonable. However, passing a structural inspection does not guarantee repository executability. In this example, the repository still fails during execution with a \texttt{ModuleNotFoundError}.

The difficulty is that this failure symptom does not by itself reveal why execution fails. The same \texttt{ModuleNotFoundError} may indicate that \emph{external dependency satisfaction} is not achieved because a required third-party package is unavailable in the current environment. It may also indicate that \emph{repository-internal reference resolution} is broken because an internal module path, import relation, or file-level reference is not correctly resolved. In other words, the same observed failure can arise from two different sources of misalignment, or from their interaction.

This ambiguity easily misleads generic generate--execute--revise pipelines. As illustrated in the upper part of Fig.~\ref{fig:motivating_example}, a baseline method can observe the failure and continue revising the repository, but the symptom alone does not indicate which problem should be addressed first. The method may keep modifying local code or import statements while the real issue is that the required external dependency is still unavailable. It may also keep revising dependency-related artifacts while the actual problem lies in broken repository-internal references. In both cases, subsequent revisions follow the visible symptom rather than the dominant source of misalignment, causing iteration to proceed in the wrong direction and preventing repository executability from being established.

In contrast, \approach does not treat the observed failure as an undifferentiated bug. Instead, it interprets execution evidence together with environment graphs to perform execution-evidence-based attribution. Specifically, \approach determines whether the dominant source of misalignment lies in external dependency satisfaction or repository-internal reference resolution, and then performs targeted repository revision accordingly. 
As illustrated in the lower part of Fig.~\ref{fig:motivating_example}, this design helps the next iteration move toward the correct revision direction rather than repeatedly reacting to surface execution symptoms.

This example highlights the central motivation of our work: in repository-level code generation, the key challenge is not merely to observe execution failure, but to identify the \emph{dominant source of misalignment} behind that failure. This observation motivates our formulation of repository-level code generation as an \emph{environment alignment} problem.

\begin{figure}[t]
    \centering
    \includegraphics[width=0.9\columnwidth]{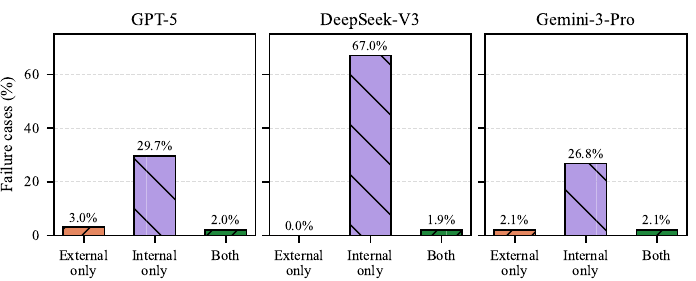}
    \caption{Environment-related failure types in failed direct generations on NL2Repo-Bench.}
    \label{fig:empirical_study}
     \vspace{-10px}
\end{figure}

\paragraph{\textbf{Motivating empirical study.}}

To examine whether failed repository-level generations already exhibit environment-related misalignment, we analyze all failed direct generations from GPT-5, DeepSeek-V3, and Gemini-3-Pro-Preview on NL2Repo-Bench \citep{ding2025nl2repo}. For each failed repository, we inspect the generated repository together with dependency manifests, import structure, build logs, runtime errors, and test outcomes. We then manually annotate whether the failure involves external dependency mismatches only, repository-internal reference resolution failures only, or both.

\textit{Main results.} Fig.~\ref{fig:empirical_study} shows that environment-related failures account for 34.7\%, 68.9\%, and 30.9\% of failed direct generations for GPT-5, DeepSeek-V3, and Gemini-3-Pro-Preview, respectively. These results directly support our motivation. In failed direct generations, environment-related failures arise not only from external dependency mismatches, but also, and more often, from repository-internal reference resolution failures. This suggests that repository executability in full-repository generation cannot be reduced to dependency installation alone. Instead, it depends on two coupled conditions: external dependency satisfaction and repository-internal reference resolution. This observation motivates our formulation of repository-level code generation as an environment alignment problem.

\section{Approach}

\subsection{Overview}

We study repository-level code generation from high-level natural-language requirements. Given a natural-language requirement $Q$, the goal is to generate a complete multi-file repository $R$ that satisfies executable validation under a target execution setting. In this setting, success is determined by repository executability of the generated repository as a whole, rather than by the local plausibility of individual files. We formulate repository executability as an environment alignment problem between the current repository and its execution environment. Successful repository execution depends on two coupled conditions: external dependency satisfaction and repository-internal reference resolution. The former concerns whether third-party dependencies, version constraints, and environment requirements can be satisfied, while the latter concerns whether files, modules, imports, and references within the repository are consistently connected and resolvable. Because failure in one condition may mask, trigger, or amplify failure in the other, the two conditions must be reasoned about jointly.

\begin{figure*}[t]
    \centering
    \includegraphics[width=0.8\textwidth]{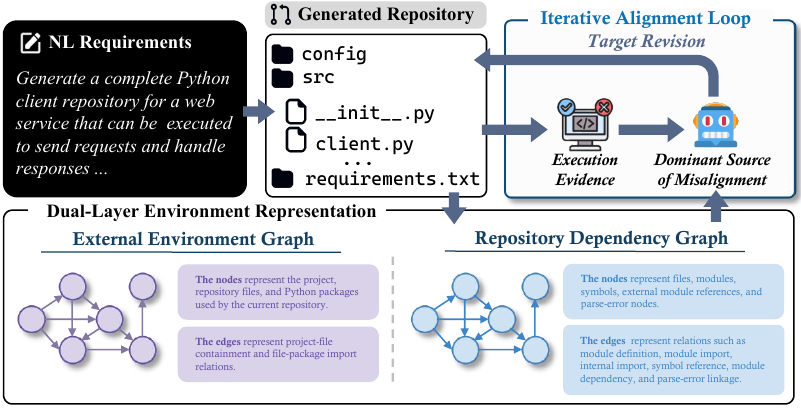}
    \caption{Overview of \approach. Starting from an initial repository, \approach builds a dual-layer environment representation, executes the repository, collects execution evidence, identifies the dominant source of misalignment, and performs targeted repository revision in an iterative alignment loop.}
    \label{fig:framework_overview}
\end{figure*}

Based on this formulation, \approach follows an iterative workflow. Starting from an initial repository $R_0$, it builds a dual-layer environment representation consisting of an external environment graph and a repository dependency graph, executes the current repository, collects execution evidence, identifies the dominant source of misalignment, and performs targeted repository revision accordingly. The attribution step considers three possible sources: unmet external dependency satisfaction, broken repository-internal reference resolution, and residual logic faults. This process repeats until the repository satisfies executable validation or the revision budget $B$ is exhausted. Figure~\ref{fig:framework_overview} illustrates the overall workflow, and Algorithm~\ref{alg:envgraph} summarizes the corresponding control logic. The following subsections describe the dual-layer environment representation, execution-evidence-based attribution, targeted repository revision, and iterative alignment loop in detail.

\begin{algorithm}[t]
\caption{Iterative environment alignment in \approach}
\label{alg:envgraph}
\begin{algorithmic}[1]
\Require high-level natural-language requirement $Q$, initial repository $R_0$, executable validation setting $\mathcal{V}$, revision budget $B$
\AlgHeader{\textbf{[Executability State]} external environment graph $G_{\mathrm{ext}}$, repository dependency graph $G_{\mathrm{int}}$}
\AlgHeader{\textbf{[Execution Evidence]} evidence set $E \gets \varnothing$}
\Ensure final repository $R_{\mathrm{final}}$

\State $R \gets R_0$

\AlgPhase{Phase I: Environment Representation Initialization}
\State $(G_{\mathrm{ext}}, G_{\mathrm{int}}) \gets \textsc{BuildEnv}(R)$

\For{$t = 1$ to $B$}

    \AlgPhase{Phase II: Execution and Evidence Collection}
    \State $E \gets \textsc{ExecRepo}(R, \mathcal{V})$

    \If{$\textsc{PassExec}(R, E, \mathcal{V})$}
        \State $R_{\mathrm{final}} \gets R$
        \State \Return $R_{\mathrm{final}}$
    \EndIf

    \AlgPhase{Phase III: Dominant Misalignment Attribution}
    \State $s \gets \textsc{Attribute}(G_{\mathrm{ext}}, G_{\mathrm{int}}, E)$
    \Statex \hspace{\algorithmicindent} $s \in \{$unmet external dependency satisfaction, broken repository-internal reference resolution, residual logic faults$\}$

    \AlgPhase{Phase IV: Targeted Repository Revision}
    \State $R \gets \textsc{Revise}(R, G_{\mathrm{ext}}, G_{\mathrm{int}}, E, s)$
    \Statex \hspace{\algorithmicindent} \textit{// a unified revision step whose focus is conditioned on $s$}

    \AlgPhase{Phase V: Executability State Update}
    \State $(G_{\mathrm{ext}}, G_{\mathrm{int}}) \gets \textsc{BuildEnv}(R)$

\EndFor

\AlgPhase{Phase VI: Final Output}
\State $R_{\mathrm{final}} \gets R$
\State \Return $R_{\mathrm{final}}$
\end{algorithmic}

\end{algorithm}

\subsection{Dual-Layer Environment Representation}

To operationalize repository executability as environment alignment, \approach maintains a dual-layer environment representation as the executability state of the current repository. This representation does not decompose repository failures into isolated subproblems. Instead, it jointly models the two coupled conditions that determine whether the repository can be successfully executed under the target validation setting: external dependency satisfaction and repository-internal reference resolution. The first layer captures whether the external packages required by the repository are available and consistently declared. The second layer captures whether repository files, modules, and symbols are consistently connected and resolvable. Together, these two layers provide the structural basis for subsequent execution-evidence-based attribution and targeted repository revision.

\subsubsection{External environment graph.}
The external environment graph, denoted by $G_{\mathrm{ext}}=(V_{\mathrm{ext}},E_{\mathrm{ext}})$, models external dependency satisfaction. In our implementation, $V_{\mathrm{ext}}$ contains a project node, repository file nodes, and external package nodes. We construct $G_{\mathrm{ext}}$ by scanning repository files, extracting import statements, and reading dependency declarations from repository manifests when available. The resulting edges encode project--file containment and file--package import relations, while each package node records whether the package is used in code and whether it is explicitly declared in the repository configuration. Through this construction, $G_{\mathrm{ext}}$ captures whether the external packages required by the current repository are available and consistently declared under the target execution setting.

\begin{figure}[t]
    \centering
    \includegraphics[width=0.8\columnwidth]{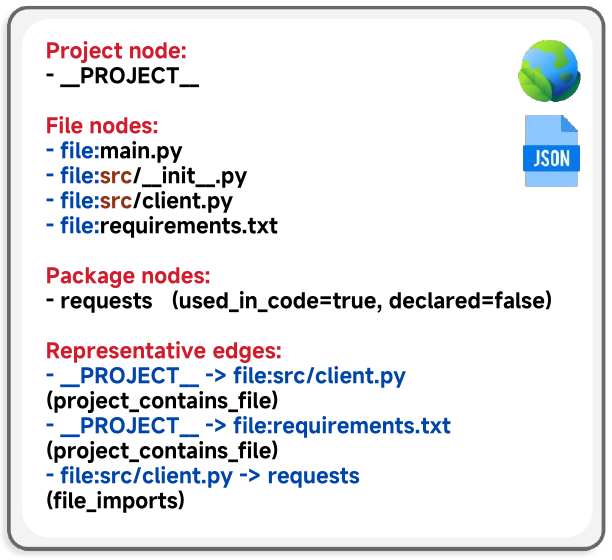}
    \caption{Example of the external environment graph for repository \texttt{Client}, showing unmet external dependency satisfaction.}
    \label{fig:env_graph_example}
\end{figure}

\textbf{\textit{An example.}} 
Figure~\ref{fig:env_graph_example} shows a concrete example based on the generated repository \texttt{Client}. The project node is connected to repository file nodes such as \texttt{main.py}, \texttt{src/\_\_init\_\_.py}, \texttt{src/client.py}, and \texttt{requirements.txt}. The file node \texttt{src/client.py} is connected to the package node \texttt{requests} through a file--package import relation. Because the package node is marked as \texttt{used\_in\_code=true} but \texttt{declared=false}, the graph directly exposes that the repository depends on an external package required by the code but missing from the repository declaration. This provides a structural explanation for the observed \texttt{ModuleNotFoundError} during execution.

\subsubsection{Repository dependency graph.}
The repository dependency graph, denoted by $G_{\mathrm{int}}=(V_{\mathrm{int}},E_{\mathrm{int}})$, models repository-internal reference resolution. In our implementation, $V_{\mathrm{int}}$ contains file nodes, module nodes, symbol nodes, unresolved module reference nodes, and parse-error nodes. We construct $G_{\mathrm{int}}$ by parsing repository files, extracting module definitions, import relations, and symbol references, and explicitly materializing unresolved imports or parse failures when they occur. The resulting edges encode file--module definition, module--module import, module--symbol definition, symbol reference, and module dependency relations. Through this construction, $G_{\mathrm{int}}$ captures whether repository files, modules, and symbols are consistently connected and resolvable during execution.

\begin{figure}[t]
    \centering
    \includegraphics[width=0.77\columnwidth]{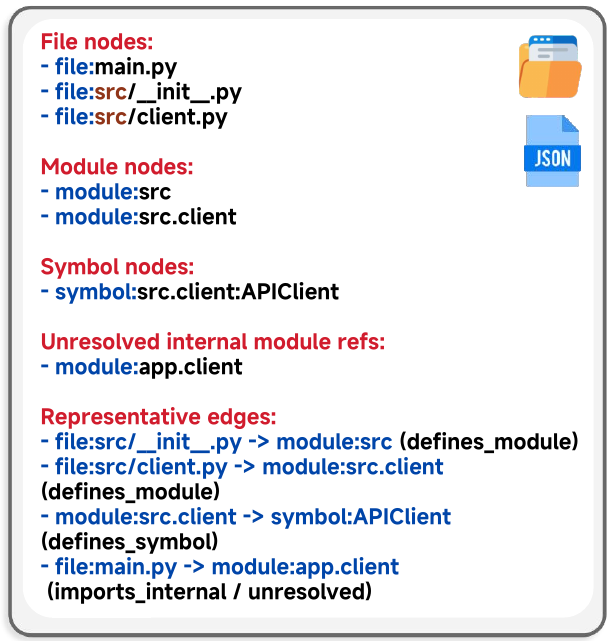}
    \caption{Example of the repository dependency graph for repository \texttt{Client}, showing broken repository-internal reference resolution caused by an unresolved internal module reference.}
    \vspace{-10px}
    \label{fig:repo_graph_example}
\end{figure}

\textbf{\textit{An example.}} 
Figure~\ref{fig:repo_graph_example} shows the corresponding repository dependency graph for the same repository. In this graph, \texttt{src/\_\_init\_\_.py} defines the internal module \texttt{src}, \texttt{src/client.py} defines the internal module \texttt{src.client}, and \texttt{src.client} further defines the symbol \texttt{APIClient}. However, \texttt{main.py} still imports the unresolved internal module reference \texttt{app.client}, which cannot be matched to any defined internal module node in the repository. As a result, the graph directly exposes broken repository-internal reference resolution, even though the repository may still appear structurally plausible at a coarse level.

These two layers must be maintained jointly. In the \texttt{Client} example, the external environment graph explains why execution fails because \texttt{requests} is required but not declared, whereas the repository dependency graph shows that execution may still fail even after the missing package is added because \texttt{main.py} refers to the unresolved internal module \texttt{app.client}. This coupling is exactly why \approach uses the two graphs jointly for subsequent execution-evidence-based attribution and targeted repository revision.

\subsection{Execution-Evidence-Based Attribution}

\textbf{\textit{Raw evidence and normalization.}}
\approach executes the current repository $R$ and collects execution evidence, including dependency installation failures, runtime errors, stack traces, and test outcomes. It then uses an LLM to normalize these heterogeneous signals, together with \(G_{\mathrm{ext}}\), \(G_{\mathrm{int}}\), and the current executability state, into a fixed evidence schema. The LLM is used only for evidence normalization, not for the final attribution decision. The full normalization schema, prompt template, and validation rules are provided in our anonymous repository.

\textbf{\textit{Attribution policy.}}
\approach applies an explicit policy over the normalized evidence to determine one dominant failure source. If the repository cannot obtain a satisfiable external execution context, the failure is attributed to failure of external dependency satisfaction. Otherwise, if unresolved internal file-, module-, or symbol-level references remain, the failure is attributed to repository-internal reference resolution failure. If neither condition holds and the repository can already be installed and launched but still fails executable validation, the failure is attributed to residual logic faults. When multiple signals co-occur, \approach resolves the ambiguity according to the prerequisite structure of executable validation: failure of external dependency satisfaction $>$ repository-internal reference resolution failure $>$ residual logic faults, since upstream external and internal failures can mask downstream logic faults.

\subsection{Targeted Repository Revision}

\textbf{\textit{Unified revision.}}
Once the dominant source of misalignment has been identified, \approach performs a unified targeted repository revision step. Rather than decomposing revision into separate repair pipelines, \approach uses a single revision mechanism whose focus is conditioned on the current executability state and the attributed source of misalignment. In this way, repository revision is not treated as a generic post-hoc correction step, but as a directed continuation of repository-level code generation toward repository executability.

\textbf{\textit{State-conditioned focus.}}
Under different dominant sources of misalignment, the same revision mechanism prioritizes different aspects of the repository. When the dominant source is unmet external dependency satisfaction, revision focuses on dependency-related inconsistencies that prevent the repository from obtaining a satisfiable external execution context. When the dominant source is broken repository-internal reference resolution, revision focuses on restoring the connectivity and resolvability of repository files, modules, and symbols. When the repository has largely achieved environment alignment but still fails executable validation, revision shifts its focus to residual logic faults and prioritizes implementation-level correction while preserving the already aligned dependency and structural conditions as much as possible.

\textbf{\textit{Graph-guided refinement.}}
The two graphs are used not only to determine the revision direction, but also to guide the revision itself. The external environment graph highlights dependency-related inconsistencies between package usage and package declaration, while the repository dependency graph highlights unresolved internal module references, broken symbol links, missing definitions, and parse failures. As a result, the same revision process is able to focus on the most critical inconsistency under the current executability state, thereby reducing ineffective corrections and more directly advancing repository executability.

\subsection{Iterative Alignment Loop}

\textbf{\textit{State-dependent process.}}
Repository executability is inherently state-dependent rather than being established in a single step. Revising one aspect of the current repository may expose previously hidden failures in another, and the dominant source of misalignment may therefore change across iterations. For this reason, \approach organizes repository-level code generation as an iterative alignment loop rather than as a one-shot revision procedure.

\textbf{\textit{Loop execution.}}
In each iteration, \approach updates the dual-layer environment representation of the current repository, executes the repository under the target validation setting, collects execution evidence, attributes the dominant source of misalignment, and then performs targeted repository revision. The revised repository then becomes the current state for the next iteration. In this way, the generation process remains grounded in the current executability state rather than in stale observations from earlier revisions.

\textbf{\textit{Stopping conditions.}}
The loop terminates in one of two cases. The first is successful termination, in which the repository satisfies executable validation and is returned as the final output. The second is budget-limited termination, in which the predefined revision budget is exhausted before repository executability is established. By repeatedly updating the repository state, execution evidence, and attributed source of misalignment together, \approach progressively advances the repository toward executability.

\section{Experiments and Results}

We evaluate \approach by defining the following research questions (RQs) and outlining how we answer them:

\begin{itemize}
    \item \textbf{RQ1. How effective is \approach for repository-level executable code generation compared with existing approaches?} We compare \approach with representative baselines across diverse backbone LLMs to assess its overall effectiveness.

    \item \textbf{RQ2. What are the individual contributions of the key components in the environment alignment design of \approach?} We conduct ablation studies on the external environment graph, the repository dependency graph, execution-evidence-based attribution, and the iterative alignment loop to examine how each component contributes to the overall effectiveness of \approach.

    \item \textbf{RQ3. What do the remaining failure cases reveal about the current limitations of \approach?} We analyze the major categories of remaining failures to understand which repository-level problems are still difficult after environment alignment.
\end{itemize}

\subsection{Experiment Settings}

\textbf{\textit{Benchmarks.}} Following prior repository-level code generation works, we conduct comprehensive experiments on two benchmarks: RAL-Bench \citep{pan2026ral} and NL2Repo-Bench \citep{ding2025nl2repo}. Both benchmarks evaluate repository-level code generation under executable validation settings, but they emphasize different aspects of the task. RAL-Bench focuses on complete repository generation from a core natural language requirement and evaluates both functional correctness and non-functional quality attributes using black-box system tests. NL2Repo-Bench focuses on long-horizon repository generation from a single requirements document and evaluates generated repositories using the upstream pytest suites of target projects. The statistics of these benchmarks are summarized in Table~\ref{tab:benchmark_stats}. In our experiments, generated repositories are executed and evaluated using the corresponding benchmark protocols and test suites.

\begin{table*}[t]
\centering
\caption{Statistics of the benchmarks used in our evaluation.}
\vspace{-5px}
\label{tab:benchmark_stats}
\small
\renewcommand{\arraystretch}{1.12}
\setlength{\tabcolsep}{4pt}
\begin{tabular}{l|c|c|c|m{8cm}}
\toprule
\textbf{Benchmark} & \textbf{\#Tasks} & \textbf{Categories} & \textbf{Input} & \textbf{Key Characteristics} \\
\hline
RAL-Bench & 38 & 7 & NL requirement & 450+ evaluation points; evaluates both functional correctness and non-functional quality \\
\hline
NL2Repo-Bench & 104 & 9 & Single requirements document & Avg. input length $\approx$ 18.8k tokens; Easy/Medium/Hard = 26/46/32 \\
\bottomrule
\end{tabular}
\end{table*}

\textbf{\textit{Metrics.}} RAL-Bench and NL2Repo-Bench use different evaluation protocols, and we therefore report benchmark-specific metrics. On RAL-Bench, we report functional correctness and non-functional quality. Functional correctness is measured by the functional test pass rate. For non-functional quality, following the ISO/IEC~25010 quality model \citep{estdale2018applying}, we aggregate five normalized dimensions, namely \textit{maintainability} (lower-bound MI from static analysis \citep{oman1992metrics,oman1994construction}), \textit{security} (high-risk static-analysis findings), \textit{robustness} (robustness-suite pass rate), \textit{efficiency} (reference-normalized runtime), and \textit{resource usage} (reference-normalized RSS memory and CPU usage), using AHP-derived weights. Formally, if \(s_i\) denotes the normalized score of the \(i\)-th dimension and \(w_i\) denotes its AHP-derived weight, the final non-functional score is computed as \(Q_{\mathrm{nf}}=\sum_i w_i s_i\), where \(\sum_i w_i = 1\). This design complements functional correctness by capturing repository quality beyond test passing alone.
On NL2Repo-Bench, we report functional correctness only, measured by the functional test pass rate under the benchmark-provided upstream pytest suites. We do not introduce an additional non-functional score on this benchmark because its protocol is designed primarily for long-horizon repository generation under executable validation. 
Detailed metric definitions, normalization formulas, and AHP derivation are provided in our repository.

\textbf{\textit{Comparative Methods.}} 
We compare \approach with representative baselines from two closely related directions: environment-aware code generation and repository-level code generation. Although these two directions partially overlap, we group baselines according to their primary modeling focus. The first direction explicitly models external execution constraints, such as library versions and API migration requirements. The second direction improves repository-level generation through planning, executable validation, and structural reasoning.

\textbf{Environment-aware code generation.}
This line explicitly incorporates environment constraints into code generation, primarily by modeling external conditions such as library versions, API compatibility, and migration requirements. We consider the following representative baselines.
\begin{itemize}
    \item \textbf{VersiCode \citep{wu2024versicode}} treats library versions as explicit generation constraints and supports version-aware code generation and editing.
    \item \textbf{APIMig \citep{kuang2025apimig}} addresses repository-level API migration across multiple library versions by combining an API evolution knowledge graph with chain exploration.
\end{itemize}

\textbf{Repository-level code generation.}
This line advances code generation beyond single functions or files by strengthening repository-level planning, executable validation, and structural reasoning. We consider the following representative baselines.
\begin{itemize}
    \item \textbf{CodePlan \citep{bairi2024codeplan}} formulates repository-level code generation as a planning problem and addresses it with a task-agnostic framework that synthesizes a multi-step chain of edits.
    \item \textbf{Repo2Run \citep{hu2025repo2run}} focuses on executable repository construction through iterative environment building, Dockerfile generation, test execution, and execution-driven output revision. Among prior repository-level methods, it is the closest baseline to our setting because it explicitly incorporates runnable-environment construction into repository-level generation.
    \item \textbf{RepoGraph \citep{ouyang2024repograph}} introduces an explicit repository-level structural representation to support repository-wide navigation and reasoning in code generation.
\end{itemize}

\textbf{\textit{Implementation details.}} We instantiate \approach with GPT-5-2025-08-07, DeepSeek-V3.2, and Gemini-3-Pro-Preview. Unless otherwise specified, we use the default context window of each model and adopt greedy decoding by setting the temperature to 0. According to prior work \citep{madaan2023self}, the maximum number of iterations is set to 4 for \approach and other iterative refinement-based baselines.

To ensure a fair comparison, all approaches are given the same problem descriptions for code generation. The generated repositories are then executed and evaluated under the corresponding benchmark protocols. This setup ensures that different approaches receive consistent external feedback, thereby enabling fair and rigorous comparison.

\subsection{RQ1: Overall Performance Comparison}

\begin{table}[t]
\centering
\caption{\textbf{Comparison of repository-level code generation methods on RAL-Bench under different backbone models.}
\textbf{Functional} and \textbf{Non-functional} denote Functional Correctness and Non-Functional Quality, respectively. Numbers are percentages. Gray numbers in parentheses indicate absolute changes relative to \textbf{Direct} under the same backbone. Red rows report its relative improvement over the best non-\approach baseline.}
\label{tab:env_method_comparison}
\small
\renewcommand{\arraystretch}{1.16}
\setlength{\arrayrulewidth}{0.8pt}
\arrayrulecolor{black}

\definecolor{DeepBlue}{RGB}{77,88,234}
\definecolor{LightBlue}{RGB}{235,241,250}
\definecolor{DeepRed}{RGB}{196,59,59}
\definecolor{LightRed}{RGB}{252,239,239}

\newcommand{\chg}[1]{\textcolor{gray}{\scriptsize\,({#1})}}
\newcommand{\best}[1]{\textcolor{DeepBlue}{\textbf{#1}}}
\newcommand{\imp}[1]{\textcolor{DeepRed}{\textbf{#1}}}

\resizebox{\columnwidth}{!}{%
\begin{tabular}{l|l|cc}
\toprule
\textbf{Model} & \textbf{Method} & \textbf{Functional} & \textbf{Non-functional} \\
\hline

\multirow{8}{*}{\textbf{GPT-5-2025-08-07}}
& \textbf{Direct}    & 42.32 & 58.26 \\
& \textbf{CodePlan}  & 52.39\chg{+10.07} & 59.01\chg{+0.75} \\
& \textbf{Repo2Run}  & 40.48\chg{-1.84}  & 56.26\chg{-2.00} \\
& \textbf{RepoGraph} & 49.33\chg{+7.01}  & 58.93\chg{+0.67} \\
& \textbf{APIMig}    & 41.31\chg{-1.01}  & 49.30\chg{-8.96} \\
& \textbf{VersiCode} & 39.37\chg{-2.95}  & 54.20\chg{-4.07} \\
& \cellcolor{LightBlue}\textcolor{DeepBlue}{\textbf{\approach}}
& \cellcolor{LightBlue}\best{55.43}\chg{+13.11}
& \cellcolor{LightBlue}\best{64.12}\chg{+5.86} \\
\hhline{~---}
& \cellcolor{LightRed}\imp{Imp.\%}
& \cellcolor{LightRed}\imp{+5.80\%}
& \cellcolor{LightRed}\imp{+8.66\%} \\
\hline

\multirow{8}{*}{\textbf{DeepSeek-V3.2}}
& \textbf{Direct}    & 27.23 & 52.16 \\
& \textbf{CodePlan}  & 30.86\chg{+3.62}  & 53.32\chg{+1.16} \\
& \textbf{Repo2Run}  & 30.12\chg{+2.89}  & 52.88\chg{+0.72} \\
& \textbf{RepoGraph} & 29.24\chg{+2.00}  & 43.10\chg{-9.06} \\
& \textbf{APIMig}    & 25.23\chg{-2.00}  & 31.52\chg{-20.64} \\
& \textbf{VersiCode} & 21.35\chg{-5.88}  & 26.74\chg{-25.42} \\
& \cellcolor{LightBlue}\textcolor{DeepBlue}{\textbf{\approach}}
& \cellcolor{LightBlue}\best{32.62}\chg{+5.39}
& \cellcolor{LightBlue}\best{55.76}\chg{+3.61} \\
\hhline{~---}
& \cellcolor{LightRed}\imp{Imp.\%}
& \cellcolor{LightRed}\imp{+5.72\%}
& \cellcolor{LightRed}\imp{+4.58\%} \\
\hline

\multirow{8}{*}{\textbf{Gemini-3-Pro-Preview}}
& \textbf{Direct}    & 38.55 & 55.10 \\
& \textbf{CodePlan}  & 42.92\chg{+4.37}  & 51.41\chg{-3.69} \\
& \textbf{Repo2Run}  & 35.80\chg{-2.75}  & 45.87\chg{-9.23} \\
& \textbf{RepoGraph} & 32.25\chg{-6.30}  & 24.78\chg{-30.32} \\
& \textbf{APIMig}    & 42.71\chg{+4.15}  & 34.38\chg{-20.72} \\
& \textbf{VersiCode} & 32.80\chg{-5.75}  & 41.24\chg{-13.86} \\
& \cellcolor{LightBlue}\textcolor{DeepBlue}{\textbf{\approach}}
& \cellcolor{LightBlue}\best{45.44}\chg{+6.89}
& \cellcolor{LightBlue}\best{59.47}\chg{+4.37} \\
\hhline{~---}
& \cellcolor{LightRed}\imp{Imp.\%}
& \cellcolor{LightRed}\imp{+5.87\%}
& \cellcolor{LightRed}\imp{+7.93\%} \\
\bottomrule
\end{tabular}%
}
\vspace{-10px}
\end{table}

\begin{table}[t]
\centering
\caption{Comparison of methods for repository-level code generation on NL2Repo-Bench under different backbone models. Imp.\% denotes relative improvement over the best non-\approach baseline.}
\label{tab:rq1_nl2repobench}
\small
\renewcommand{\arraystretch}{1.1}
\setlength{\arrayrulewidth}{0.8pt}
\arrayrulecolor{black}

\definecolor{DeepBlue}{RGB}{77,88,234}
\definecolor{LightBlue}{RGB}{235,241,250}
\definecolor{DeepRed}{RGB}{196,59,59}
\definecolor{LightRed}{RGB}{252,239,239}

\newcommand{\chg}[1]{\textcolor{gray}{\scriptsize\,({#1})}}
\newcommand{\best}[1]{\textcolor{DeepBlue}{\textbf{#1}}}
\newcommand{\imp}[1]{\textcolor{DeepRed}{\textbf{#1}}}

\resizebox{\columnwidth}{!}{%
\begin{tabular}{l|cccc}
\toprule
\multirow{2}{*}{\textbf{Method}} & \textbf{Overall} & \textbf{Easy} & \textbf{Medium} & \textbf{Hard} \\
 & \textbf{Score (\%)} & ($\leq$1.5k LOC) & (1.5k--4k LOC) & ($\geq$4k LOC) \\
\midrule

\multicolumn{5}{c}{\textcolor{DeepBlue}{\textbf{\textit{GPT-5-2025-08-07}}}} \\
\midrule
\textbf{Direct}      & 21.7 & 38.4 & 20.7 & 9.6 \\
\textbf{CodePlan}    & 30.0\chg{+8.3} & 50.2\chg{+11.8} & 29.6\chg{+8.9} & 14.2\chg{+4.6} \\
\textbf{Repo2Run}    & 30.0\chg{+8.3} & 50.4\chg{+12.0} & 29.5\chg{+8.8} & 14.2\chg{+4.6} \\
\textbf{RepoGraph}   & 29.2\chg{+7.5} & 51.3\chg{+12.9} & 28.0\chg{+7.3} & 12.9\chg{+3.3} \\
\textbf{APIMig}      & 30.6\chg{+8.9} & 47.2\chg{+8.8} & 30.8\chg{+10.1} & 17.0\chg{+7.4} \\
\textbf{VersiCode}   & 28.0\chg{+6.3} & 47.3\chg{+8.9} & 26.2\chg{+5.5} & 15.0\chg{+5.4} \\
\rowcolor{LightBlue}
\textcolor{DeepBlue}{\textbf{\approach}} 
& \cellcolor{LightBlue}\best{33.2}\chg{+11.5}
& \cellcolor{LightBlue}\best{48.0}\chg{+9.6}
& \cellcolor{LightBlue}\best{33.2}\chg{+12.5}
& \cellcolor{LightBlue}\best{21.3}\chg{+11.7} \\
\hhline{-----}
\rowcolor{LightRed}
\imp{Imp.\%}         & \imp{+8.4\%} & \imp{-6.4\%} & \imp{+7.8\%} & \imp{+25.3\%} \\
\midrule

\multicolumn{5}{c}{\textcolor{DeepBlue}{\textbf{\textit{DeepSeek-V3.2}}}} \\
\midrule
\textbf{Direct}      & 25.0 & 33.6 & 22.8 & 8.8 \\
\textbf{CodePlan}    & 23.3\chg{-1.8} & 32.1\chg{-1.5} & 22.4\chg{-0.4} & 7.1\chg{-1.7} \\
\textbf{Repo2Run}    & 28.6\chg{+3.6} & 39.8\chg{+6.2} & 27.5\chg{+4.7} & 8.8\chg{+0.0} \\
\textbf{RepoGraph}   & 23.3\chg{-1.7} & 32.1\chg{-1.5} & 22.5\chg{-0.3} & 7.9\chg{-0.9} \\
\textbf{APIMig}      & 22.3\chg{-2.7} & 32.0\chg{-1.6} & 21.6\chg{-1.2} & 5.9\chg{-2.9} \\
\textbf{VersiCode}   & 25.8\chg{+0.8} & 33.6\chg{+0.0} & 23.7\chg{+0.9} & 9.5\chg{+0.7} \\
\rowcolor{LightBlue}
\textcolor{DeepBlue}{\textbf{\approach}} 
& \cellcolor{LightBlue}\best{28.9}\chg{+3.9}
& \cellcolor{LightBlue}\best{33.2}\chg{-0.4}
& \cellcolor{LightBlue}\best{29.9}\chg{+7.1}
& \cellcolor{LightBlue}\best{16.1}\chg{+7.3} \\
\hhline{-----}
\rowcolor{LightRed}
\imp{Imp.\%}         & \imp{+1.0\%} & \imp{-16.6\%} & \imp{+8.7\%} & \imp{+69.5\%} \\
\midrule

\multicolumn{5}{c}{\textcolor{DeepBlue}{\textbf{\textit{Gemini-3-Pro-Preview}}}} \\
\midrule
\textbf{Direct}      & 34.2 & 44.9 & 40.9 & 16.8 \\
\textbf{CodePlan}    & 42.0\chg{+7.8} & 50.8\chg{+5.9} & 46.6\chg{+5.7} & 28.2\chg{+11.4} \\
\textbf{Repo2Run}    & 43.0\chg{+8.8} & 51.0\chg{+6.1} & 46.9\chg{+6.0} & 30.3\chg{+13.5} \\
\textbf{RepoGraph}   & 42.2\chg{+8.0} & 49.9\chg{+5.0} & 45.8\chg{+4.9} & 29.1\chg{+12.3} \\
\textbf{APIMig}      & 41.6\chg{+7.4} & 48.5\chg{+3.6} & 45.4\chg{+4.5} & 28.6\chg{+11.8} \\
\textbf{VersiCode}   & 37.1\chg{+2.9} & 42.6\chg{-2.3} & 41.5\chg{+0.6} & 21.8\chg{+5.0} \\
\rowcolor{LightBlue}
\textcolor{DeepBlue}{\textbf{\approach}} 
& \cellcolor{LightBlue}\best{44.02}\chg{+5.72}
& \cellcolor{LightBlue}\best{50.8}\chg{+6.6}
& \cellcolor{LightBlue}\best{46.3}\chg{+5.4}
& \cellcolor{LightBlue}\best{35.2}\chg{+18.4} \\
\hhline{-----}
\rowcolor{LightRed}
\imp{Imp.\%}         & \imp{+5.72\%} & \imp{+14.7\%} & \imp{+13.2\%} & \imp{+52.5\%} \\
\bottomrule
\end{tabular}%
}
\vspace{-10px}
\end{table}

The main results on RAL-Bench are reported in Table~\ref{tab:env_method_comparison}. Across all three backbone models, \approach consistently achieves the best performance on both Functional Correctness and Non-Functional Quality. Compared with Direct, \approach improves Functional Correctness by 13.11, 5.39, and 6.89 points on GPT-5-2025-08-07, DeepSeek-V3.2, and Gemini-3-Pro-Preview, respectively. It also improves Non-Functional Quality by 5.86, 3.61, and 4.37 points under the three backbones. Compared with the strongest non-\approach baseline, which is CodePlan under all three backbones, \approach still achieves consistent relative gains, reaching 5.80\%--5.87\% in Functional Correctness and 4.58\%--8.66\% in Non-Functional Quality. These results show a consistent advantage of \approach across the tested backbone models. The NL2Repo-Bench results, as shown in Table~\ref{tab:rq1_nl2repobench}, follow a consistent pattern. \approach achieves the highest overall score across all settings, outperforming other methods in the overall score.

\textbf{\textit{Comparison with Repository-Level Baselines.}}
A closer comparison with repository-level baselines further highlights the advantages of \approach. Methods such as CodePlan, Repo2Run, and RepoGraph provide partial improvements in some settings, suggesting that planning, executable validation, and structural reasoning can benefit repository-level code generation. Among them, CodePlan is the strongest non-\approach baseline across all three backbones. However, these gains remain limited and are not always consistent on Non-Functional Quality, as illustrated by the performance drop of CodePlan on Gemini-3-Pro-Preview. This suggests that repository-level mechanisms alone can improve task-relevant generation, but they are still insufficient to reliably ensure repository executability together with broader engineering quality.

\textbf{\textit{Comparison with Environment-Aware Baselines.}}
The comparison with environment-aware baselines reveals a different limitation. APIMig and VersiCode explicitly model external execution constraints, such as API evolution, dependency migration, and version adaptation. However, their performance remains noticeably weaker and less stable across backbone models, and several results are even worse than Direct, especially on Non-Functional Quality. This pattern suggests that modeling only external execution conditions is insufficient for repository-level code generation, where successful generation also depends on repository-internal reference resolution and repository-wide structural consistency.

\textbf{\textit{Cross-Backbone Trends.}}
Another notable observation is that the relative ranking of baselines varies substantially across backbone models, whereas \approach remains consistently strongest. This indicates that repository-level code generation is sensitive to the capability profile of the underlying model, and that many existing baselines do not maintain stable gains across different backbones. In contrast, \approach delivers consistent improvements under all three tested models, suggesting that its design captures a more reliable mechanism for improving repository executability across diverse backbone settings.

\begin{tcolorbox}[
  enhanced,
  colback=grey,
  colframe=teal!60!black,
  boxrule=0.6pt,
  arc=10pt,
  left=4mm,right=4mm,
  top=2mm,bottom=2mm,
  drop shadow={black!40!white},
]
\textbf{\textit{Answer to RQ1:} }
\textit{\approach achieves the best results across all compared methods and backbone models on both Functional Correctness and Non-Functional Quality. These findings show that jointly addressing external dependency satisfaction and repository-internal reference resolution is critical for repository executability, enabling \approach to maintain a consistent advantage over representative repository-level and environment-aware baselines.}
\end{tcolorbox}

\subsection{RQ2: Ablation Study}

\begin{table}[t]
\centering
\caption{Ablation study of \approach on RAL-Bench under different backbone models. EEBA denotes execution-evidence-based attribution. Gray numbers in parentheses indicate absolute drops relative to the full model under the same backbone.}
\label{tab:ablation_envgraph}
\small
\renewcommand{\arraystretch}{1.12}
\setlength{\tabcolsep}{4pt}

\definecolor{DeepBlue}{RGB}{77,88,234}
\definecolor{LightBlue}{RGB}{235,241,250}
\newcommand{\chg}[1]{\textcolor{gray}{\scriptsize\,({#1})}}
\newcommand{\best}[1]{\textcolor{DeepBlue}{\textbf{#1}}}

\begin{tabular}{c|l|cc}
\toprule
\textbf{Model} & \textbf{Variant} & \textbf{Func.} & \textbf{Non-Func.} \\
\hline

\multirow{5}{*}{\centering\arraybackslash\shortstack{GPT-5\\-2025-08-07}}
& \cellcolor{LightBlue}\textcolor{DeepBlue}{\textbf{\approach}} 
& \cellcolor{LightBlue}\best{55.43} 
& \cellcolor{LightBlue}\best{64.12} \\
& w/o External Env. Graph & 52.57\chg{-2.86} & 59.92\chg{-4.20} \\
& w/o Repo. Dep. Graph & 48.44\chg{-6.99} & 62.76\chg{-1.36} \\
& w/o EEBA & 44.07\chg{-11.36} & 58.56\chg{-5.56} \\
& w/o Iterative Alignment Loop & 43.33\chg{-12.10} & 57.97\chg{-6.15} \\
\hline

\multirow{5}{*}{\centering\arraybackslash\shortstack{DeepSeek\\-V3.2}}
& \cellcolor{LightBlue}\textcolor{DeepBlue}{\textbf{\approach}} 
& \cellcolor{LightBlue}\best{32.62} 
& \cellcolor{LightBlue}\best{55.76} \\
& w/o External Env. Graph & 29.33\chg{-3.29} & 54.73\chg{-1.03} \\
& w/o Repo. Dep. Graph & 31.56\chg{-1.06} & 55.03\chg{-0.73} \\
& w/o EEBA & 31.96\chg{-0.66} & 55.46\chg{-0.30} \\
& w/o Iterative Alignment Loop & 31.56\chg{-1.06} & 55.29\chg{-0.47} \\
\hline

\multirow{5}{*}{\centering\arraybackslash\shortstack{Gemini-3\\Pro-Preview}}
& \cellcolor{LightBlue}\textcolor{DeepBlue}{\textbf{\approach}} 
& \cellcolor{LightBlue}\best{45.44} 
& \cellcolor{LightBlue}\best{59.47} \\
& w/o External Env. Graph & 43.17\chg{-2.27} & 58.75\chg{-0.72} \\
& w/o Repo. Dep. Graph & 43.29\chg{-2.15} & 58.53\chg{-0.94} \\
& w/o EEBA & 42.41\chg{-3.03} & 57.58\chg{-1.89} \\
& w/o Iterative Alignment Loop & 42.41\chg{-3.03} & 57.28\chg{-2.19} \\
\bottomrule
\end{tabular}
\end{table}

The ablation results are presented in Table~\ref{tab:ablation_envgraph}. Across all three backbone models, the full \approach consistently achieves the best performance on both Functional Correctness and Non-Functional Quality. Removing any component leads to performance degradation, indicating that all four components make positive contributions to repository-level code generation and repository executability.

\textbf{\textit{Effect of Environment Graphs.}}
Both graph modules are beneficial, but they contribute in different ways. Removing the external environment graph consistently reduces both Functional Correctness and Non-Functional Quality, highlighting the importance of modeling dependency availability and version compatibility. Removing the repository dependency graph causes a larger decline in Functional Correctness, especially on GPT-5 (-6.99), suggesting that this module is particularly important for repository-internal reference resolution. These results indicate that the two graphs provide complementary support for external dependency satisfaction and repository-internal reference resolution.

\textbf{\textit{Effect of Execution-Evidence-Based Attribution and the Iterative Alignment Loop.}}
Execution-evidence-based attribution and the iterative alignment loop have the largest impact, particularly on GPT-5. Removing execution-evidence-based attribution decreases Functional Correctness and Non-Functional Quality by 11.36 and 5.56 points, respectively, while removing the iterative alignment loop causes even larger drops of 12.10 and 6.15 points. These findings show that environment modeling alone is insufficient; the framework must also identify the dominant source of misalignment and perform targeted repository revision based on execution evidence.

\textbf{\textit{Cross-Backbone Trends.}}
The overall trend is consistent across backbone models, although the magnitudes of the drops vary. The effects are strongest on GPT-5, moderate on Gemini-3-Pro-Preview, and smaller on DeepSeek-V3.2, suggesting that stronger backbones are better able to exploit structured environment signals and iterative alignment. Nevertheless, all ablated variants perform worse than the full \approach on every backbone, confirming the consistent advantage of the complete design across the evaluated backbones.

\begin{tcolorbox}[
  enhanced,
  colback=grey,
  colframe=teal!60!black,
  boxrule=0.6pt,
  arc=10pt,
  left=4mm,right=4mm,
  top=2mm,bottom=2mm,
  drop shadow={black!40!white},
]
\textbf{\textit{Answer to RQ2:} }
\textit{All four components contribute to the final performance of \approach. Among them, execution-evidence-based attribution and the iterative alignment loop have the largest impact, while the external environment graph and the repository dependency graph provide complementary support for external dependency satisfaction and repository-internal reference resolution.}
\end{tcolorbox}

\subsection{RQ3: Error Analysis}

\begin{figure*}[t]
    \centering
    \subfloat[\textit{GPT-5-2025-08-07}]{
        \includegraphics[width=0.22\textwidth]{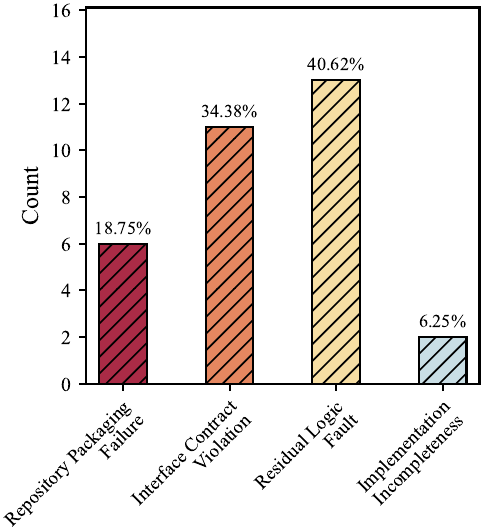}
        \label{fig:rq4_gpt5}
    }\hspace{0.01\textwidth}
    \subfloat[\textit{DeepSeek-V3.2}]{
        \includegraphics[width=0.22\textwidth]{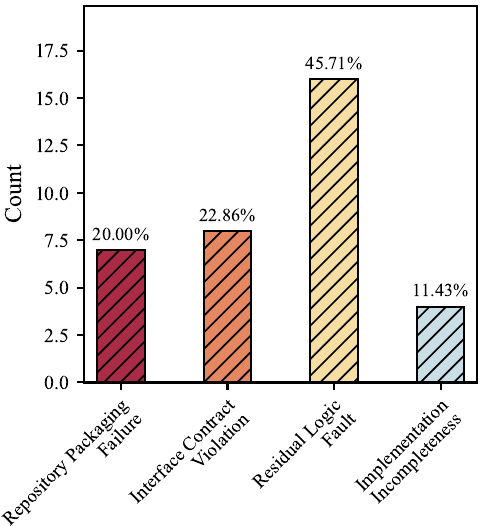}
        \label{fig:rq4_deepseek}
    }\hspace{0.01\textwidth}
    \subfloat[\textit{Gemini-3-Pro-Preview}]{
        \includegraphics[width=0.22\textwidth]{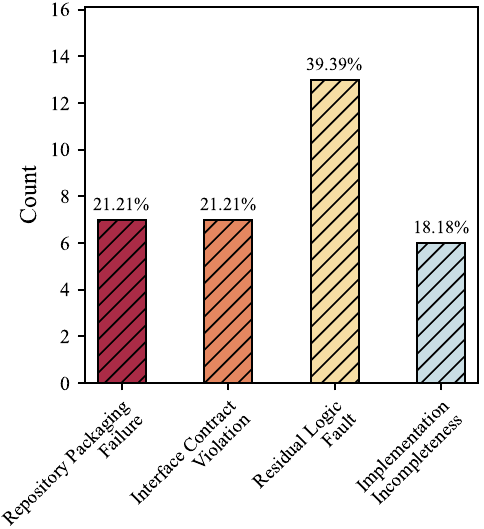}
        \label{fig:rq4_gemini}
    }
    \caption{Failure-type distribution of \approach across different backbone models.}
    \label{fig:rq4_failure_analysis}
\end{figure*}

For this RQ, we analyze the remaining failures of \approach using GPT-5-2025-08-07, DeepSeek-V3.2, and Gemini-3-Pro-Preview on RAL-Bench. Figure~\ref{fig:rq4_failure_analysis} shows the distribution of major failure types across the three backbone models. This analysis aims to reveal which repository-level problems remain difficult after environment alignment and therefore still limit repository executability.

\textbf{\textit{Failure Assignment.}} We manually inspect the remaining failed repositories produced by \approach on RAL-Bench for each backbone model. We assign one primary failure category to each failed case based on the generated repository, execution logs, stack traces, and test outcomes. When multiple issues co-occur, we use the dominant failure type as the assigned label.

\textbf{\textit{Failure Modes.}} We use the following failure categories in this analysis:


\begin{itemize}
    \item \textcolor{fm1}{\textbf{Repository Packaging Failures:}} Failures caused by improper repository organization, broken packaging structure, or missing files that prevent successful repository execution. These failures are closely related to unresolved repository-level structural and reference issues.
    \item \textcolor{fm2}{\textbf{Interface Contract Violations:}} Failures caused by inconsistent function signatures, incorrect parameter passing, or mismatched module-level invocation assumptions across repository components.
    \item \textcolor{fm3}{\textbf{Residual Logic Faults:}} Failures in which the repository can be executed but still does not implement the required functionality correctly.
    \item \textcolor{fm4}{\textbf{Implementation Incompleteness:}} Failures caused by missing, unfinished, or only partially implemented code that prevents full repository functionality.
\end{itemize}

\textbf{\textit{Results.}} Residual Logic Faults account for the largest share of the remaining failures across all three backbone models. For GPT-5-2025-08-07, Residual Logic Faults account for 40.62\% of all analyzed failures. For DeepSeek-V3.2, this category further rises to 45.71\%. For Gemini-3-Pro-Preview, Residual Logic Faults remain the largest category at 39.39\%. This pattern suggests that environment alignment alone is not sufficient for full repository executability, and that final success still depends heavily on accurate logic generation.

At the same time, the distributions also reveal model-specific differences in the remaining bottlenecks. DeepSeek-V3.2 exhibits a relatively higher proportion of Repository Packaging Failures (20.00\%), suggesting that repository-level structural construction remains a nontrivial challenge for this model. Gemini-3-Pro-Preview shows a more balanced failure distribution, with Interface Contract Violations (21.21\%) and Implementation Incompleteness (18.18\%) also contributing substantially. This suggests that, beyond logic correctness, cross-module interface consistency and complete implementation remain important challenges in repository-level code generation.

Overall, the remaining failures are distributed across logic, structure, interface consistency, and completeness, rather than being dominated by a single non-logic factor. This indicates that \approach alleviates part of the environment-related bottleneck, but full repository executability still depends on stronger end-to-end implementation quality and cross-file coordination.

\begin{tcolorbox}[
  enhanced,
  colback=grey,                
  colframe=teal!60!black,       
  boxrule=0.6pt,                
  arc=10pt,                     
  left=4mm,right=4mm,           
  top=2mm,bottom=2mm,
  drop shadow={black!40!white}, 
]
\textbf{\textit{Answer to RQ3:} }
\textit{The remaining failures show that environment alignment is necessary but not sufficient for full repository executability. Residual Logic Faults are the main remaining failure source, while repository structure, interface coordination, and implementation completeness also deserve attention. In addition to improving alignment, stronger end-to-end implementation and cross-file coordination are promising directions for future work.}

\end{tcolorbox}


\section{Threats to Validity}

\textbf{Threats in generalizability.}
Although we evaluate \approach using three diverse backbone models and compare it against representative repository-level and environment-aware baselines, the current evaluation still cannot cover the full diversity of repository-level code generation scenarios. Our experiments also consider both functional correctness and non-functional quality, which reduces the risk that our conclusions depend on a single metric. However, the benchmark scope, software stacks, and execution environments considered in this work remain limited. Therefore, our findings should be interpreted as evidence of effectiveness within the evaluated settings rather than as proof of universal generalizability. In future work, we plan to further examine the generalizability of \approach on broader benchmark collections, more diverse software stacks, and additional backbone models.

\textbf{Threats in benchmark contamination.}
Since modern LLMs are trained on large-scale corpora, it is difficult to completely rule out the possibility that some benchmark repositories, dependency specifications, or related implementation patterns may have appeared in the training data. Such contamination may influence the absolute level of performance achieved by all methods. However, this threat applies to all compared methods under the same backbone model. Therefore, although benchmark contamination may affect absolute scores, it does not invalidate the fairness of our comparative analysis or the relative improvements of \approach, which remain consistent across different backbone settings.

\textbf{Threats in method scope.}
Our framework focuses on failures related to external dependency satisfaction and repository-internal reference resolution, and uses execution evidence to identify the dominant source of misalignment and guide targeted repository revision. Although this design is effective for the failure space considered in our evaluation, it may not fully cover all execution failures encountered in real-world repositories. For example, some failures may arise from hidden infrastructure assumptions, undocumented build or deployment logic, dynamically generated resources, or external services that are not explicitly modeled in the current framework. When failures are dominated by such factors, the attribution and revision process may become less accurate. We therefore view \approach as a principled step toward environment-aligned repository-level code generation rather than as a complete solution to every real-world execution failure mode. Extending the framework to broader classes of environment and infrastructure constraints is an important direction for future work.

\section{Related Work}

\textbf{Environment-Aware Code Generation.}
Early studies on LLM-based code generation are typically conducted under standardized evaluation settings in which the runtime environment is fixed in advance \citep{chen2021evaluating, joel2024survey, 11242137}. Under this assumption, models are expected to generate code without explicit access to library versions, API compatibility requirements, or other execution constraints, and success is evaluated primarily based on functional correctness \citep{yu2024codereval, sen2025large, wu2024comprehensive}. Subsequent studies have shown that this assumption is often too restrictive: code that appears functionally correct may still fail in the target environment because of unmet dependency requirements, version incompatibility, or API evolution \citep{hu2025repo2run, wu2026environment, cheng2025codemenv}. This observation has motivated a growing body of research on environment-aware code generation, which explicitly models execution constraints rather than treating them as implicit assumptions \citep{islah2024gitchameleon, wang2025llms, kuhar2025libevolutioneval, wu2024versicode}. Representative efforts in this direction primarily focus on modeling external environment conditions. VersiCode \citep{wu2024versicode} explicitly incorporates library version constraints into code generation, treating version compatibility as a first-class objective rather than an implicit prerequisite. Extending this perspective beyond local code generation, APIMig \citep{kuang2025apimig} addresses repository-level API migration with the support of an API evolution knowledge graph, enabling migration across multiple library versions rather than isolated API replacement. Overall, these studies establish the importance of environment constraints in code generation and substantially improve generation under changing dependency and API conditions. However, they mainly formulate environment awareness in terms of external dependency satisfaction, such as library versions, API compatibility, and migration constraints, without jointly modeling repository-internal reference resolution as another necessary condition for repository executability.

\textbf{Repository-Level Code Generation.}
A parallel line of research moves beyond single-function or single-file generation and investigates repository-level code generation in multi-file software projects \citep{zhang2024systematic, hu2025assessing, yang2025code, li2024deveval, jimenez2023swe, le2025impacts, pan2025codecorllmbasedselfreflectivemultiagent}. CodePlan \citep{bairi2024codeplan} shows that repository-level generation cannot be reduced to local retrieval alone and instead formulates code generation as a planning problem that coordinates edits across files and components. Repo2Run \citep{hu2025repo2run} further shifts the focus from producing plausible code to improving repository executability by iteratively building runnable environments and revising outputs based on execution feedback. Complementarily, RepoGraph \citep{ouyang2024repograph} introduces an explicit graph representation of repository structure and dependencies, enabling more accurate cross-file reasoning and refinement. Taken together, these methods substantially advance repository-level code generation in terms of planning, executable validation, and structural reasoning \citep{bairi2024codeplan}. However, they are still primarily designed around repository coordination, execution-driven revision, or structural dependency modeling, rather than formulating repository executability itself as a unified environment alignment problem. In particular, they do not explicitly determine whether an observed execution failure is mainly caused by unmet external dependency satisfaction, broken repository-internal reference resolution, or residual logic faults after the environment is largely aligned \citep{hu2025repo2run}.

\textbf{Our Position.}
Prior work has advanced either the modeling of external execution constraints or repository-level planning, executable validation, and structural reasoning, but has rarely unified them as a single problem of environment alignment for repository executability \citep{wu2024versicode, kuhar2025libevolutioneval, kuang2025apimig, bairi2024codeplan, hu2025repo2run}. Our core idea is to jointly model two coupled conditions for repository executability: external dependency satisfaction and repository-internal reference resolution. Based on this unified perspective, \approach uses execution evidence to perform execution-evidence-based attribution, identifies the dominant source of misalignment, and then performs targeted repository revision, thereby forming an iterative alignment loop for repository-level code generation and refinement.

\section{Conclusion}

In this paper, we propose \approach, a framework for repository-level code generation that formulates repository executability as an environment alignment problem. It jointly models external dependency satisfaction and repository-internal reference resolution through a dual-layer environment representation, and improves repository generation via execution-evidence-based attribution, unified targeted repository revision, and an iterative alignment loop. Extensive experiments on diverse benchmarks and LLMs demonstrate the superior performance of \approach. Our work represents a promising step toward environment-aligned repository-level code generation and executable repository construction. In future work, we plan to evaluate \approach on broader benchmark collections and additional backbone models. We will also explore how the environment alignment framework can be extended to other repository-scale software engineering tasks beyond repository-level code generation. 



\bibliographystyle{ACM-Reference-Format}
\bibliography{references}

\end{document}